\documentclass[fleqn,usenatbib,useAMS]{mnras}
 \usepackage{graphicx}
 \usepackage{amsmath}
 \usepackage{amssymb}
 \usepackage[T1]{fontenc}
 \usepackage{ae,aecompl}
 \usepackage{txfonts}
 \usepackage{enumerate}

 \title[The analemma criterion for quasi-satellites]
       {The analemma criterion: accidental quasi-satellites are indeed true 
        quasi-satellites}

 \author[C. de la Fuente Marcos and R. de la Fuente Marcos]
        {C.~de~la~Fuente~Marcos\thanks{E-mail: carlosdlfmarcos@gmail.com}
         and
         R. de la Fuente Marcos \\
         Apartado de Correos 3413, E-28080 Madrid, Spain}
 \date{Accepted 2016 July 22.
       Received 2016 July 22;
       in original form 2016 July 1}
 \pubyear{2016}
 
\begin{document}
  \label{firstpage}
  \pagerange{\pageref{firstpage}--\pageref{lastpage}}
  \maketitle

  \begin{abstract}
     In the Solar system, a quasi-satellite is an object that follows a 
     heliocentric path with an orbital period that matches almost exactly
     with that of a host body (planetary or not). The trajectory is of such 
     nature that, without being gravitationally attached, the value of the 
     angular separation between host and quasi-satellite as seen from the 
     Sun remains confined within relatively narrow limits for time-spans 
     that exceed the length of the host's sidereal orbital period. Here, we 
     show that under these conditions, a quasi-satellite traces an analemma 
     in the sky as observed from the host in a manner similar to that found 
     for geosynchronous orbits. The analemmatic curve (figure-eight-, 
     teardrop-, ellipse-shaped) results from the interplay between the tilt 
     of the rotational axis of the host and the properties of the orbit of 
     the quasi-satellite. The analemma criterion can be applied to identify 
     true quasi-satellite dynamical behaviour using observational or 
     synthetic astrometry and it is tested for several well-documented 
     quasi-satellites. For the particular case of 15810 (1994~JR$_1$), a 
     putative accidental quasi-satellite of dwarf planet Pluto, we show 
     explicitly that this object describes a complex analemmatic curve for 
     several Plutonian sidereal periods, confirming its transient 
     quasi-satellite status. 
  \end{abstract}


  \section{Introduction}
     Objects trapped in a 1:1 mean motion resonance with a host (planetary or not) are classified as co-orbitals of the host, independently
     of the shape and orientation of their paths (Morais \& Morbidelli 2002); in other words, to be classed as co-orbitals their orbits do 
     not have to resemble that of the host as long as the ratio of their orbital periods equates to almost exactly one. In general, 
     co-orbital configurations are not identified observationally but as a result of the statistical analysis of large sets of numerical 
     integrations. There is, however, a potential exception to this standard approach; a particular type of co-orbital configuration that 
     can be confirmed observationally, the quasi-satellite dynamical state.

     Here, we study the apparent motion in host-centric equatorial coordinates of known quasi-satellites to show that they trace an 
     analemmatic curve in the sky as observed from the host in a manner similar to that found for geosynchronous orbits. This paper is 
     organized as follows. Section~2 discusses the so-called analemma criterion for quasi-satellites and it includes an extensive 
     exploration of the known quasi-satellite population. The particular case of 15810 (1994~JR$_1$), a putative accidental quasi-satellite 
     of Pluto, is analysed in Section~3 to show that according to the analemma criterion it is a true transient quasi-satellite of Pluto. 
     Results are discussed in Section~4 and conclusions are summarized in Section~5. 

  \section{The analemma criterion}
     Minor bodies are confirmed as co-orbitals after statistical analysis of their simulated orbital evolution. Here, we show that there is
     a particular type of co-orbital configuration that may be confirmed observationally: the quasi-satellite state.
   
     \subsection{Quasi-satellites: a short review and a lost specimen}
        Minor bodies engaged in quasi-satellite behaviour with a host move near the host for the duration of the quasi-satellite episode
        although each pair minor-body--host is not gravitationally bound. In the Solar system and from a frame of reference centred at the 
        Sun but corotating with the host, the quasi-satellite appears to go around the host like a regular retrograde satellite but the 
        physical distance between the two bodies is always greater than the radius of the Hill sphere of the host (see e.g. fig. 1 in 
        Mikkola et al. 2004). The quasi-satellite state is one of the dynamical epitomes of the 1:1 mean motion or co-orbital resonance, the 
        other two being the Trojan or tadpole and the horseshoe resonant states (see e.g. Murray \& Dermott 1999; Mikkola et al. 2006).

        Dynamical classification within the 1:1 mean motion resonance is based on the study of a critical angle, the relative mean 
        longitude, $\lambda_{\rm r}$, or difference between the mean longitude of the object and that of its host. If $\lambda_{\rm r}$ 
        librates or oscillates over time, then the object under study is a co-orbital. In principle, this can only be confirmed via $N$-body 
        simulations. Quasi-satellites exhibit libration of $\lambda_{\rm r}$ about 0\degr (for additional details, see e.g. Mikkola et al. 
        2006; de la Fuente Marcos \& de la Fuente Marcos 2014, 2016a,b). 

        The existence of quasi-satellites was predicted more than a century ago (Jackson 1913), but the first bona fide quasi-satellite
        was not identified until much later ---2002~VE$_{68}$ was confirmed as quasi-satellite of Venus by Mikkola et al. (2004). However,
        the first quasi-satellite may have been identified in 1973 although it was apparently lost shortly after. Using numerical 
        integrations, Chebotarev (1974)\footnote{Originally published in Russian, Astron. Zh., 50, 1071-1075 (1973 September--October).} 
        showed that the so-called minor planet 7617 (see his fig. 5 and table 5) was a quasi-satellite of Jupiter although he regarded this 
        object as a distant Jovian satellite. This minor planet 7617 is clearly (see table 5 in Chebotarev 1974) not asteroid 7617 
        (1996~VF$_{30}$), as Chebotarev (1974) followed the numbering scheme in van Houten et al. (1970). The orbital elements (1950 
        equinox) of the mysterious minor planet 7617 in van Houten et al. (1970) ---$a=5.0785$~au, $e=0.6179$, $i=4\fdg080$, 
        $\Omega=68\fdg81$ and $\omega=209\fdg23$--- do not match those of any known asteroid or comet; therefore, it is pressumed lost.
        
     \subsection{Theoretical expectations}
        When observed from a celestial object (planetary or not) true satellites (not following synchronous orbits), co-orbitals of the 
        Trojan or horseshoe type, and passing objects describe roughly sinusoidal paths in the sky over a sidereal orbital period. In sharp 
        contrast, quasi-satellites appear to orbit the host when viewed in a heliocentric frame of reference that rotates with the host. As 
        their orbital periods are very close to the sidereal period of the host, the standard sinusoidal trace becomes compressed 
        longitudinally turning into an analemmatic curve. 

        The analemma or analemmatic curve ---the figure-eight loop--- has been traditionally linked to graphic depictions of the changing of 
        the seasons and the equation of time (see e.g. Heath 1923; Raisz 1942; Oliver 1972; di Cicco 1979; Irvine 2001; Holbrow 2013). In 
        addition, the trajectories of geosynchronous satellites as observed from the ground have the appearance of an analemma (Chalmers 
        1987). From the host, the apparent motion of a quasi-satellite during a sidereal orbital period is not too different from that of 
        true satellites moving in synchronous orbits.

        In order to show that a given quasi-satellite traces an analemmatic curve, we proceed as follows. We perform full $N$-body 
        simulations in ecliptic coordinates; at time $t$, for a given host of coordinates, $(x_{\rm h}, y_{\rm h}, z_{\rm h})$, and axial 
        tilt or obliquity, $\epsilon$, and a certain quasi-satellite located at $(x_{\rm qs}, y_{\rm qs}, z_{\rm qs})$ we can define the
        host-centric equatorial coordinates, $(\alpha^{*}, \delta^{*})$:
        \begin{equation}
           \begin{aligned}
              r \cos\alpha^{*}\ \cos\delta^{*} & = x_{\rm qs} - x_{\rm h} \\
              r \sin\alpha^{*}\ \cos\delta^{*} & = (y_{\rm qs} - y_{\rm h})\cos\epsilon - (z_{\rm qs} - z_{\rm h})\sin\epsilon \\
              r \sin\delta^{*}                 & = (y_{\rm qs} - y_{\rm h})\sin\epsilon + (z_{\rm qs} - z_{\rm h})\cos\epsilon \,, 
              \label{eqnconv}
           \end{aligned}
        \end{equation}
        where $r$ is the distance between host and quasi-satellite at time $t$. For the particular case of the Earth, $(\alpha^{*}, 
        \delta^{*})$ become $(\alpha, \delta)$, the usual geocentric equatorial coordinates. Over one sidereal period, there is a 
        north-south oscillation of $\delta^{*}$ that is responsible for the lengthwise extension of the analemma pattern. Such libration is 
        induced by the fact that the orbital plane of the quasi-satellite and the celestial equator at the host are, in general, tilted by a 
        certain amount. In addition, the relative motion of a quasi-satellite with respect to its host is not uniform because, in a typical 
        case, their orbital eccentricities are different although their semimajor axes are nearly equal and this tends to distort the 
        analemma. The observed apparent motion results from the interplay between the two effects; when both have comparable strengths, the 
        familiar figure-eight is obtained. In the particular case of the Earth and for an ideal quasi-satellite bright enough to be observed 
        year-round with standard ground-based telescopes, regular astrometric observations should make it possible to plot the associated 
        analemmatic curve without any help from numerical computations. Unfortunately, no such quasi-satellite is known to exist and 
        $N$-body simulations are needed to produce synthetic astrometry to confirm the theoretical expectations.

\begin{figure*}
      \centering
       \includegraphics[width=0.49\linewidth]{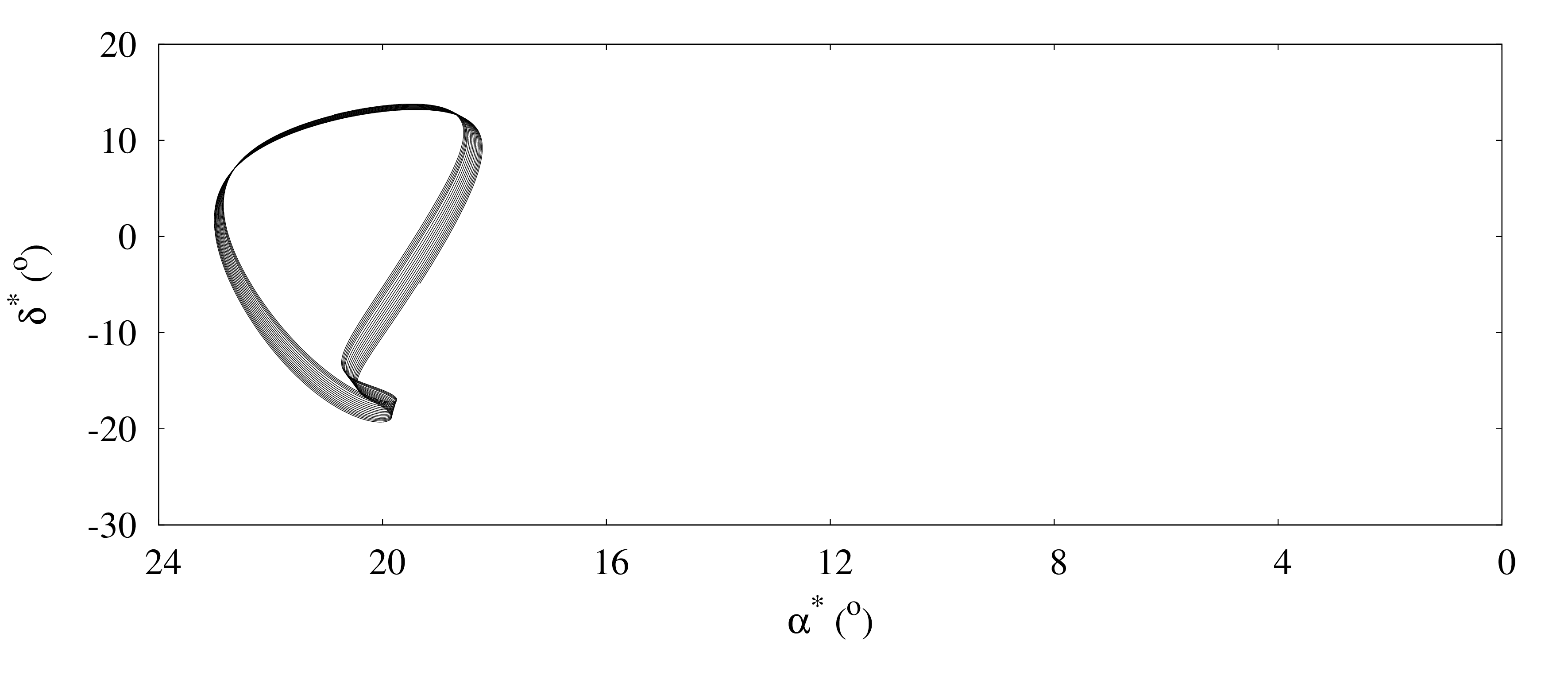}
       \includegraphics[width=0.49\linewidth]{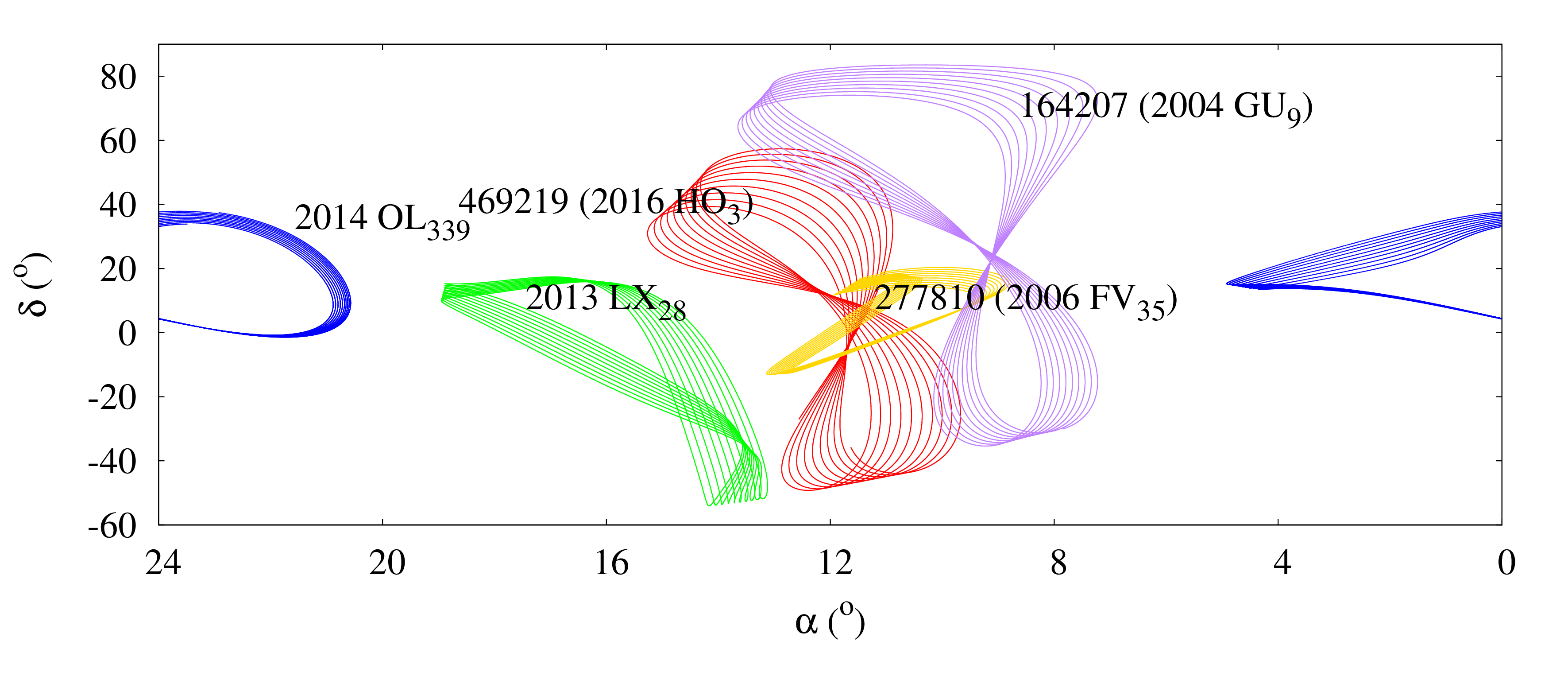} \\
       \includegraphics[width=0.49\linewidth]{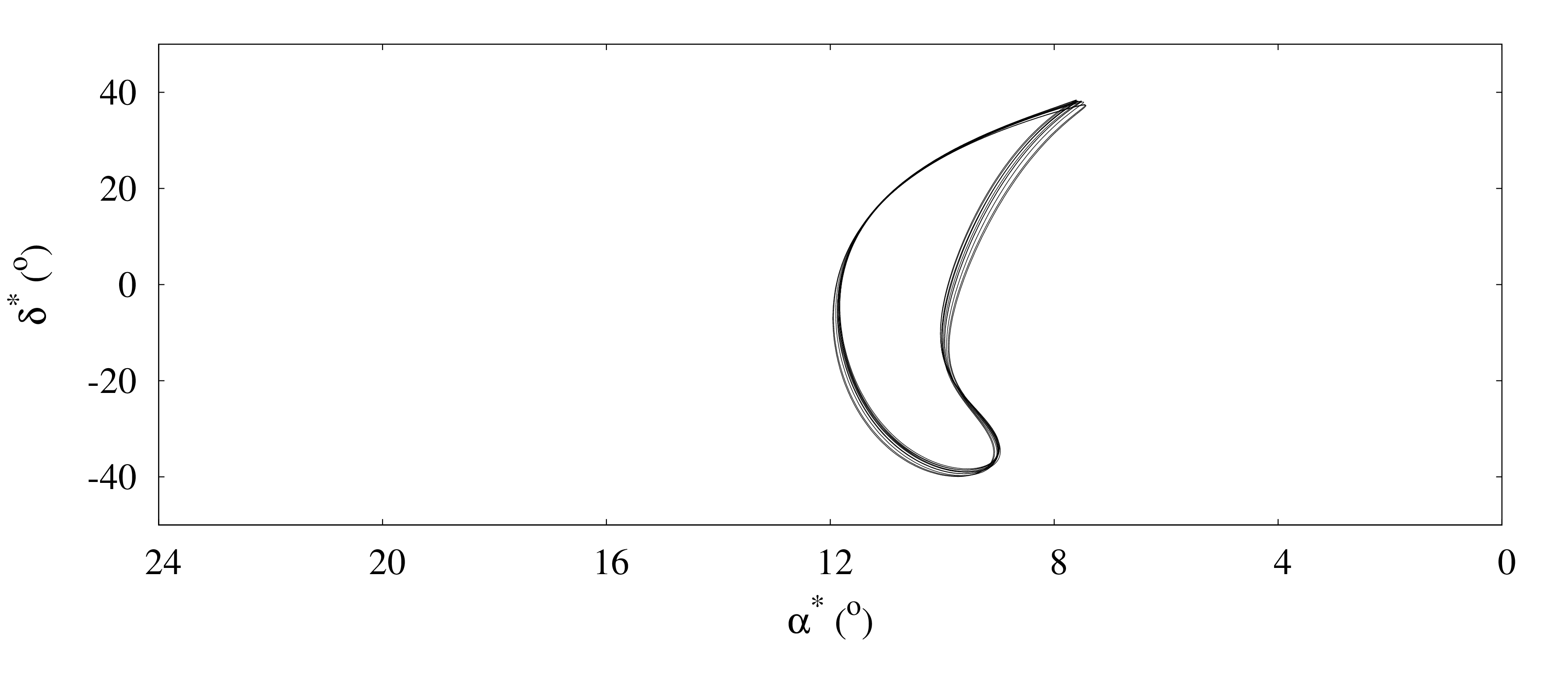}
       \includegraphics[width=0.49\linewidth]{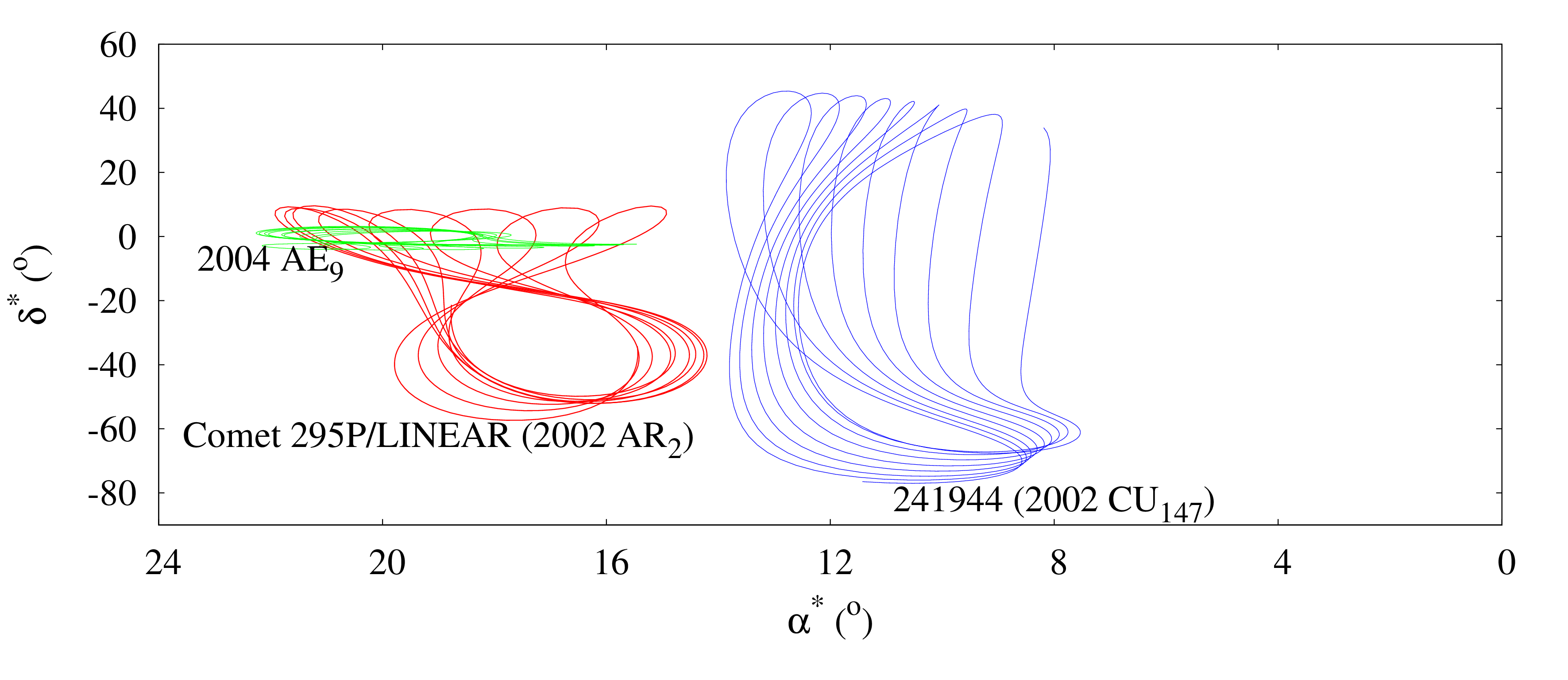} \\
       \includegraphics[width=0.49\linewidth]{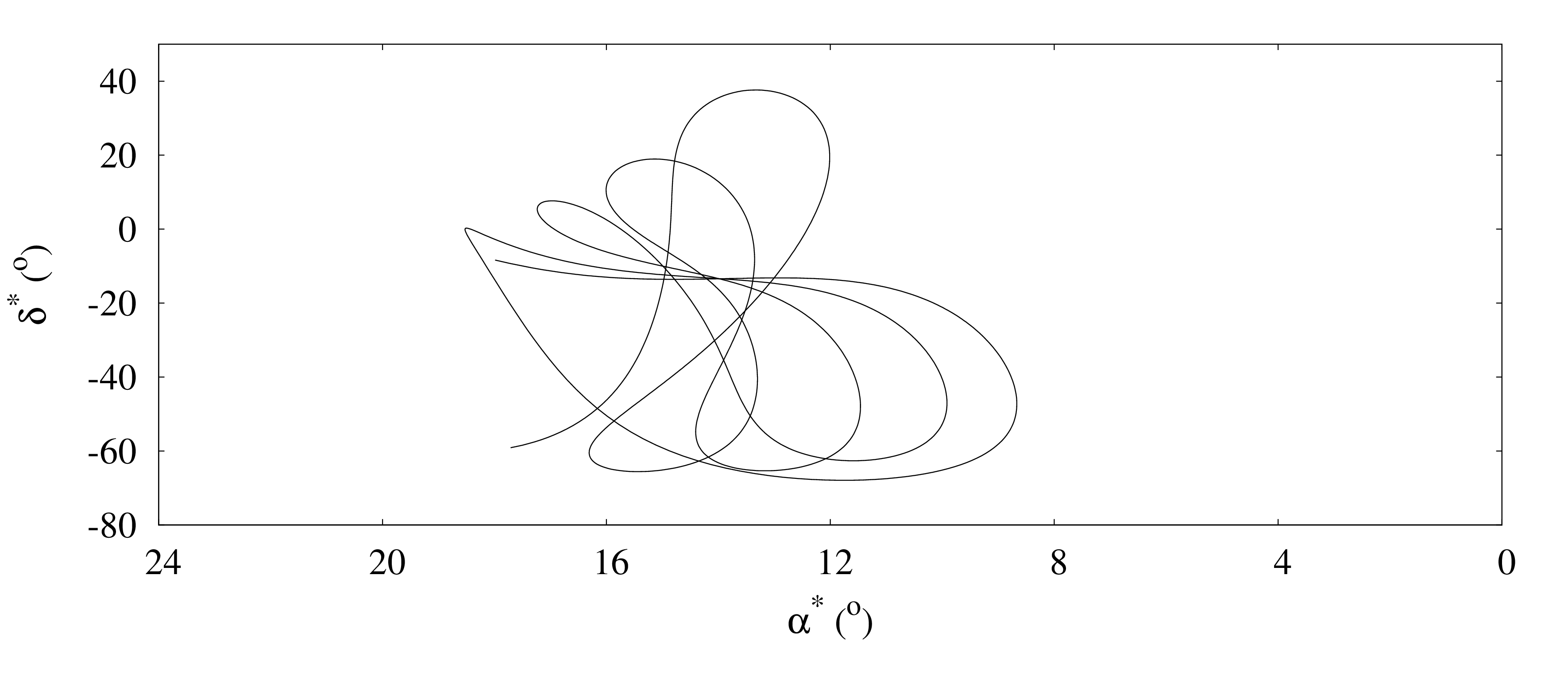}
       \includegraphics[width=0.49\linewidth]{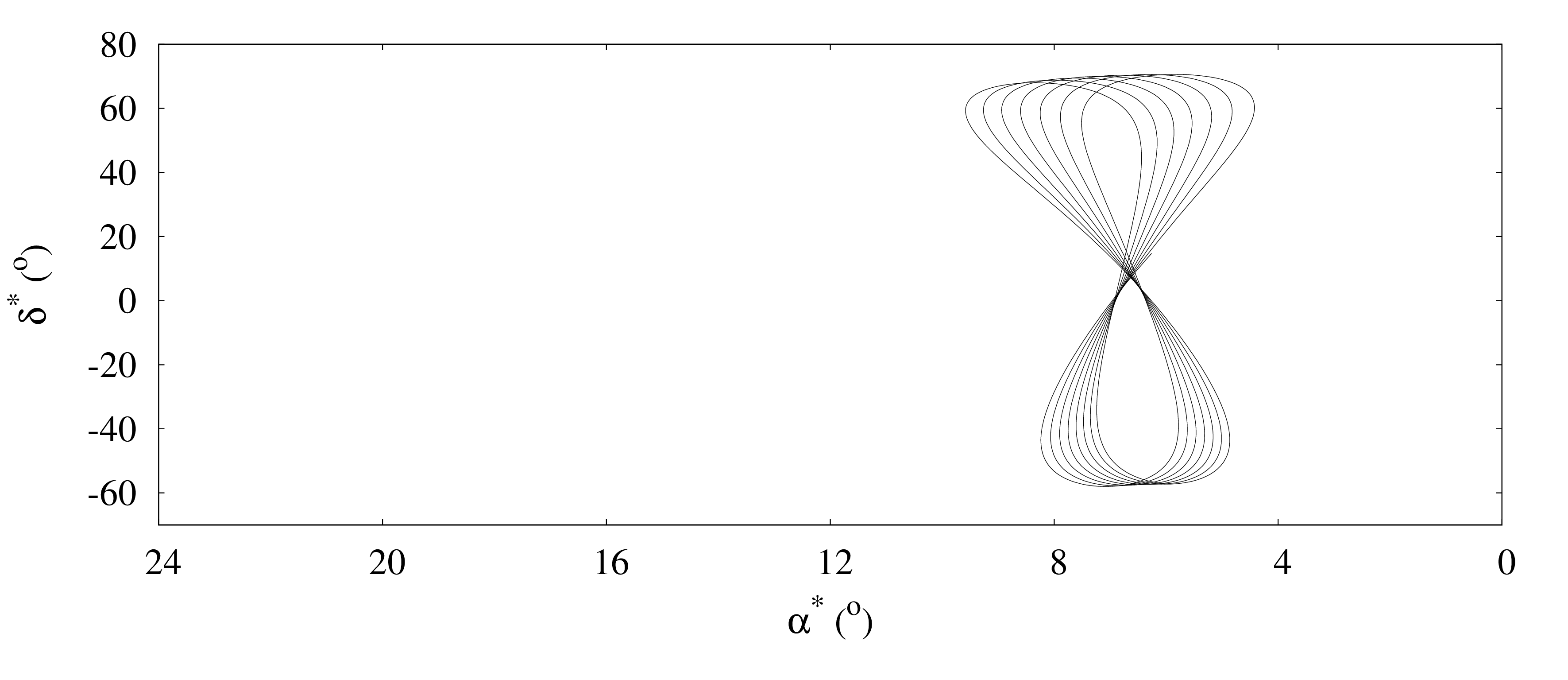} \\
       \caption{Apparent motion in host-centric equatorial coordinates of known quasi-satellites. Asteroid 2002~VE$_{68}$ from Venus (top
                left-hand panel), 164207 (2004~GU$_{9}$), 277810 (2006~FV$_{35}$), 2013~LX$_{28}$, 2014~OL$_{339}$ and 469219 
                (2016~HO$_{3}$) from the Earth (top right-hand panel), 76146 (2000~EU$_{16}$) from Ceres (middle left-hand panel), comet 
                295P/LINEAR (2002~AR$_{2}$), 241944 (2002~CU$_{147}$) and 2004~AE$_{9}$ from Jupiter (middle right-hand panel), 63252 
                (2001~BL$_{41}$) from Saturn (bottom left-hand panel), and 309239 (2007~RW$_{10}$) from Neptune (bottom right-hand panel).
               }
       \label{qs}
    \end{figure*}

        The apparent motion of the objects studied here and plotted in Figs \ref{qs} and \ref{Pqs} has been computed using the Hermite 
        scheme (Makino 1991; Aarseth 2003).\footnote{http://www.ast.cam.ac.uk/$\sim$sverre/web/pages/nbody.htm} The Cartesian state vectors 
        of the integrated bodies at the epoch 2457600.5 JD TDB (2016-July-31.0) have been retrieved from the Jet Propulsion Laboratory's 
        (JPL) \textsc{horizons}\footnote{http://ssd.jpl.nasa.gov/?horizons} system (Giorgini et al. 1996); this epoch is the $t$ = 0 instant 
        in the simulations. Full details of these calculations can be found in de la Fuente Marcos \& de la Fuente Marcos (2012b, 2014, 
        2016a,b). 
%
%
%
%

     \subsection{The case of Venus}
        Asteroid 2002~VE$_{68}$ was confirmed as quasi-satellite of Venus by Mikkola et al. (2004). The orbital evolution of this object was 
        further studied in de la Fuente Marcos \& de la Fuente Marcos (2012b). Its orbit is quite eccentric ($e=0.4103$) and moderately 
        inclined ($i=9\fdg0070$). Fig. \ref{qs}, top left-hand panel, shows the results of equations (\ref{eqnconv}) from $t=0$ until 10 yr 
        later, i.e. over 16 orbital sidereal periods of Venus. Over 16 analemmatic loops are displayed and, consistent with its significant 
        eccentricity, one of the lobes of the analemma is very small, each loop resembling an inverted teardrop.  

     \subsection{The case of the Earth}
        Our calculations show that our planet hosts the largest known population of quasi-satellites in the Solar system; however, their 
        dynamical origin appears to be rather heterogeneous (de la Fuente Marcos \& de la Fuente Marcos 2014, 2016a,b). There are five 
        confirmed quasi-satellites of the Earth: 164207 (2004~GU$_{9}$) (Connors et al. 2004; Mikkola et al. 2006; Wajer 2010), 277810 
        (2006~FV$_{35}$) (Wiegert et al. 2008; Wajer 2010), 2013~LX$_{28}$ (Connors 2014), 2014~OL$_{339}$ (de la Fuente Marcos \& de la 
        Fuente Marcos 2014, 2016a) and 469219 (2016~HO$_{3}$)\footnote{http://www.jpl.nasa.gov/news/news.php?feature=6537} (de la Fuente 
        Marcos \& de la Fuente Marcos 2016b). Fig. \ref{qs}, top right-hand panel, shows the apparent motion of these five objects over 10 
        sidereal periods; a wide range of behaviours, from a very symmetric figure-eight to very distorted teardrop shapes, is observed.
 
        Asteroid 164207 (purple) has both moderate eccentricity ($e=0.1362$) and inclination ($i=13\fdg6491$), and it traces a somewhat 
        symmetric figure-eight that slowly shifts, keeping the position of the node almost fixed. Asteroid 277810 (gold) has significant
        eccentricity ($e=0.3776$), but low orbital inclination ($i=7\fdg1041$); consistently, its apparent motion describes a very distorted
        teardrop as the effect of the eccentricity dominates. Asteroid 2013~LX$_{28}$ (green) follows a quite eccentric ($e=0.4520$) and 
        very inclined ($i=49\fdg9754$) path that translates into an apparent motion that traces an elongated teardrop-shaped curve. Asteroid
        2014~OL$_{339}$ (blue) describes a somewhat elliptic analemma which suggests that one of the effects is nearly negligible; 
        consistently, it follows a very eccentric ($e=0.4608$) but moderately inclined ($i=10\fdg1868$) orbit. Finally, the orbit pursued by 
        469219 (red) has both low eccentricity ($e=0.1041$) and inclination ($i=7\fdg77140$); consistently, the analemma described in the 
        sky resembles that of 164207 with a rather symmetric shifting figure-eight.  

     \subsection{The case of Ceres}
        Planets are not the only possible hosts of quasi-satellite bodies, dwarf planet Ceres also has one of these interesting companions 
        (Christou 2000; Christou \& Wiegert 2012), 76146 (2000~EU$_{16}$). Fig. \ref{qs}, middle left-hand panel, shows the results of 
        nearly 11 sidereal periods. Asteroid 76146 follows a low-eccentricity ($e=0.1674$), low-inclination ($i=8\fdg8475$) orbit; 
        surprisingly, its relatively low eccentricity is high enough to induce a rather distorted teardrop shape to the resulting analemma.

     \subsection{The case of Jupiter}
        Jupiter is often regarded as the host of the largest known population of quasi-satellites with at least six, including asteroids and 
        comets (Kinoshita \& Nakai 2007; Wajer \& Kr\'olikowska 2012). However, we failed to confirm several of the proposed candidates as
        present-day quasi-satellites of Jupiter. We have found three confirmed quasi-satellites of Jupiter: comet 295P/LINEAR (2002~AR$_{2}$) 
        (Kinoshita \& Nakai 2007; Wajer \& Kr\'olikowska 2012), 241944 (2002~CU$_{147}$) (Wajer \& Kr\'olikowska 2012) and 2004~AE$_{9}$ 
        (Kinoshita \& Nakai 2007; Wajer \& Kr\'olikowska 2012). Fig. \ref{qs}, middle right-hand panel, shows the results of 8.4 sidereal 
        periods of apparent motion for these objects. Comet 295P/LINEAR (red) follows a very eccentric orbit ($e=0.6460$) that is
        only moderately inclined ($i=14\fdg6912$); consistently, a rather asymmetric figure-eight is observed. This object is unlikely to be
        the mysterious minor planet 7617 in van Houten et al. (1970), the angular elements being too different. Asteroid 241944 (blue) 
        pursues a relatively eccentric orbit ($e=0.3136$) that is also quite inclined ($i=32\fdg8906$); the relatively rapidly shifting 
        analemma exhibits somewhat symmetric lobes. In sharp contrast, 2004~AE$_{9}$ follows a very eccentric ($e=0.6459$) but nearly 
        ecliptic ($i=1\fdg6521$) orbit and its apparent motion as seen from Jupiter traces a very squashed teardrop. Again, the angular 
        elements of 2004~AE$_{9}$ are very different from those of the mysterious minor planet 7617.  

     \subsection{The case of Saturn}
        Gallardo (2006) indicated that 15504 (1999~RG$_{33}$) could be a quasi-satellite of Saturn. Our calculations show that it is indeed 
        a transient co-orbital of Saturn, but not a quasi-satellite like the objects previously discussed. Its apparent motion somewhat 
        resembles that of Molniya-type artificial satellites of the Earth (the apocentre of the very eccentric orbits occurs at a large 
        declination). However, 63252 (2001~BL$_{41}$) that was discovered by Gehrels et al. (2001) is currently a short-lived 
        quasi-satellite of Saturn that will change its current dynamical status in about 130 yr from now. Asteroid 63252 follows an 
        eccentric ($e=0.2948$) but moderately inclined ($i=12\fdg5163$) orbit. Fig. \ref{qs}, bottom left-hand panel, shows the apparent 
        motion of this object as seen from Saturn for about 4.4 sidereal periods, prior to leaving its current quasi-satellite state. The 
        analemmatic loops described by this object are very distorted as a result of its unstable dynamical behaviour and the lobes have 
        somewhat different sizes because the effect derived from eccentricity is stronger than that of inclination. Although its orbital 
        evolution is rather chaotic, 63252 ---an organic rich D-type asteroid (Doressoundiram et al. 2003)--- has been pre-selected by NASA 
        for an in situ exploration mission (Ryan et al. 2009). 

     \subsection{The case of Neptune}
        Asteroid 309239 (2007~RW$_{10}$) is so far the only confirmed quasi-satellite of Neptune (de la Fuente Marcos \& de la Fuente Marcos 
        2012a) and it is one of the largest known co-orbital companions in the Solar system with a diameter of about 250 km. It follows an 
        eccentric ($e=0.3004$) and rather inclined ($i=36\fdg1755$) orbit. Fig. \ref{qs}, bottom right-hand panel, shows 7.3 sidereal 
        periods of a very regular analemma of the figure-eight type with both lobes of nearly the same size which confirms that the effects 
        derived from eccentricity and inclination have very similar strength in this case. 

  \section{Accidental quasi-satellites: the case of Plutino 15810 (1994 JR$_1$)}
     Quasi-satellites are not exclusive of planetary hosts as the case of 76146 (2000~EU$_{16}$) and Ceres confirms. Yu \& Tremaine (1999) 
     and Tiscareno \& Malhotra (2009) used numerical simulations to predict the existence of minor bodies experiencing quasi-satellite 
     behaviour with respect to dwarf planet Pluto. Plutino 15810 (1994 JR$_1$) was identified as an accidental quasi-satellite of Pluto 
     by de la Fuente Marcos \& de la Fuente Marcos (2012c). It was termed accidental because, for this object, $\lambda_{\rm r}$ circulates 
     with a superimposed libration resulting from the oscillation of the orbital period induced by the 2:3 mean motion resonance with 
     Neptune. Such libration plays a role in triggering and terminating the quasi-satellite phase. Porter et al. (2016) have used astrometry 
     acquired by NASA's New Horizons spacecraft to improve the already robust orbital solution available for this object (see Appendix A) 
     and revisit its quasi-satellite status. The new data have been used to argue that the quasi-satellite nature of 15810 must be 
     rejected.\footnote{http://www.nasa.gov/feature/new-horizons-collects-first-science-on-a-post-pluto-object}

\begin{figure}
      \centering
       \includegraphics[width=\linewidth]{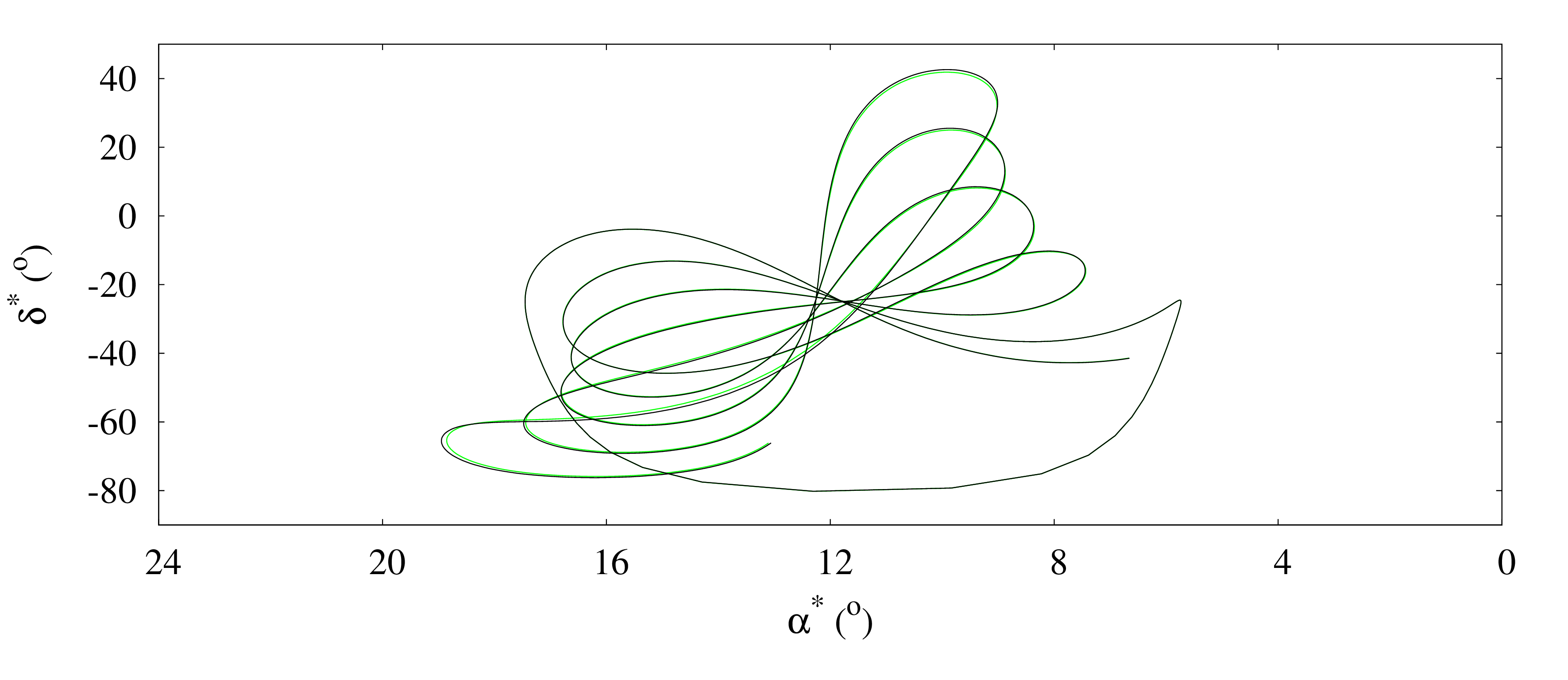} 
       \caption{Apparent motion in Pluto-centric equatorial coordinates of 15810 (1994~JR$_1$). The black curve shows the shifting analemma
                resulting from the dynamical evolution of the latest version of the orbit that includes data from NASA's New Horizons 
                spacecraft (third orbital solution in Table \ref{orb}). The green curve is equivalent to the black one but using the second 
                orbital solution in Table \ref{orb}.
               }
       \label{Pqs}
    \end{figure}

     Fig. \ref{Pqs} clearly shows that although the orbital solution of 15810 has been indeed greatly improved using New Horizons data (see 
     Appendix A), its orbital evolution still matches the one described in de la Fuente Marcos \& de la Fuente Marcos (2012c). In black, we
     have the apparent motion resulting from the latest orbit available for 15810 (third orbital solution in Table \ref{orb}). The figure
     displays the time interval of interest ---the one showing analemmatic behaviour--- that goes from 1200 years prior to $t=0$ to 200 
     years afterwards or about 5.6 sidereal orbital periods of Pluto. The analemma shifts rapidly and it is quite distorted because the 
     orbits of both Pluto and 15810 are eccentric and there is a chaotic interaction between the two bodies. Within the context of the 
     analemma criterion, the behaviour observed in Fig. \ref{Pqs} is not very different from that of some of the objects in Fig. \ref{qs}. 
     The apparent motion of 15810 closely resembles that of comet 295P/LINEAR (2002~AR$_{2}$) or 63252 (2001~BL$_{41}$). If the pre-NASA's 
     New Horizons distant encounter orbit (second orbital solution in Table \ref{orb}) is used (green curve), the differences are minimal. 
     Porter et al. (2016) argue that, for 15810, instead of transient quasi-satellite behaviour we should speak of periodic (every 2.4 Myr)
     scattering conjunctions; however, Fig. \ref{Pqs} suggests that 15810 is not different from comet 295P or 63252 in dynamical terms when 
     the analemma criterion is applied during one of its encounters with Pluto. Therefore, Pluto has at least one present-day (transient and 
     recurrent) quasi-satellite, 15810. 
%
%
%
%

  \section{Discussion}
     Our analysis so far argues in favour of a conjecture: objects that follow a quasi-satellite path with respect to a host trace an 
     analemmatic curve in the sky as observed from the host over a sidereal orbital period. Conversely, any object tracing an analemmatic 
     curve in the sky as observed from the host over a sidereal year must be a quasi-satellite of the host. Unfortunately, a rigorous 
     mathematical or even a numerical proof of the assumed theorem and its inverse is out of the scope of this work because (1) a complete
     theory of quasi-satellite motion still remains elusive (see e.g. Mikkola et al. 2006) and (2) the relevant volume of the orbital 
     parameter space to be explored is simply too large. Intuitively, the truthfulness of our conjecture can hardly be argued. The position
     of the quasi-satellite as seen from the host is subjected to two periodic librations. At some point during the sidereal year, the 
     quasi-satellite is east of the host, very nearly half an orbital period later it is west from the host. As the quasi-satellite 
     bean-shaped loop (see e.g. fig. 1 in Mikkola et al. 2004) drifts back and forth, the peri-host shifts from eastwards to westwards from 
     the host (this causes the loop drift). This behaviour is mainly the result of the difference in eccentricity between host and 
     quasi-satellite and drives the oscillation in host-centric right ascension. The relative inclination between the equatorial plane of 
     the host and the orbital plane of the quasi-satellite drives the oscillation in host-centric declination. These two oscillations have 
     (nearly) commensurable frequencies because host and quasi-satellite have very similar orbital periods and generate the analemma.

     Co-orbitals have been traditionally classified as such after the statistical analysis of numerical simulations. However, an algorithm 
     to decide whether an object is a quasi-satellite of a given host, not based on $N$-body simulations, is described in detail in section 
     4 of Mikkola et al. (2006). The analemmas or analemmata in Figs \ref{qs} and \ref{Pqs} show that, in the particular case of 
     quasi-satellites, astrometry can be readily used to perform a reliable classification. Figs \ref{qs} and \ref{Pqs} also show that both 
     very regular ---164207 (2004~GU$_{9}$), 469219 (2016~HO$_{3}$) or 309239 (2007~RW$_{10}$)--- and rather irregular ---comet 295P/LINEAR 
     (2002~AR$_{2}$), 63252 (2001~BL$_{41}$) or 15810 (1994~JR$_1$)--- short-term evolutions are possible. The analemma traced by the 
     quasi-satellite encodes relevant orbital information. Distorted, rapidly shifting analemmas are characteristic of quasi-satellites 
     moving in strongly perturbed orbits. 

     Our calculations show that, if a suitable quasi-satellite is found, it can be used as a permanent platform to install instrumentation 
     that may be used to monitor permanently the host body and enable a relatively stable communications relay for subsequent missions (e.
     g. landing quasi-autonomous vehicles on the host) at zero fuel cost because the quasi-satellite behaves like a geosynchronous satellite 
     from the point of view of the host body. For this task, the smaller its average distance from the host the better (see e.g. the case of 
     469219 as discussed in de la Fuente Marcos \& de la Fuente Marcos 2016b). Artificial quasi-satellites are also possible (see e.g. Kogan 
     1989 for the Phobos mission). For quasi-satellites sufficiently close to a host, substantial parallax may occur; therefore, and 
     depending on the location of the observer on the surface of the host, different analemmas may be observed. This issue together with the 
     shift of the analemma loop induced by orbital evolution requires robotic tracking of the quasi-satellite yearly movement around its 
     analemma. The use of quasi-satellite trajectories in astrodynamics has been frequently discussed (see e.g. Kogan 1989; Lidov \& 
     Vashkov'yak 1993, 1994; Mikkola \& Prioroc 2016). It could be the case that the analemmatic behaviour described here had been found 
     before within the context of astrodynamical studies (see e.g. Kogan 1989).

  \section{Conclusions}
     In this paper, we have explored a new criterion to identify quasi-satellites. In sharp contrast with the numerical strategies 
     customarily applied in the study of co-orbital bodies, the criterion described here can make direct use of observational astrometric 
     data. Our conclusions can be summarized as follows.
     \begin{enumerate}[(i)]
        \item Bona fide quasi-satellites trace paths in the sky which repeat every sidereal period when observed from their hosts. These
              paths can be described as analemmatic curves similar to those found for geosynchronous orbits. The analemma shifts as the 
              orbit of the quasi-satellite changes over time. 
        \item The existence of this analemmatic behaviour turns quasi-satellites, natural or artificial, into potentially interesting 
              platforms for the future of space exploration.
        \item The Earth has the largest known number of present-day quasi-satellites, five. Jupiter comes in second place with three.
              Venus, Saturn, Neptune and dwarf planet Ceres have one each. 
        \item Applying the analemma criterion, Plutino 15810 (1994 JR$_1$) is as good a quasi-satellite as it may get. Therefore, dwarf 
              planet Pluto hosts at least one quasi-satellite at present.
        \item Asteroid 63252 (2001~BL$_{41}$) is a present-day transient quasi-satellite of Saturn.
        \item Historically, the first object identified as quasi-satellite (in this case of Jupiter) was an asteroid moving in a comet-like 
              orbit. Unfortunately, this object appears to have been lost since its announcement back in 1973.  
     \end{enumerate}

  \section*{Acknowledgements}
     We thank the referee, S. Mikkola, for his prompt reports, S.~J. Aarseth for providing the code used in this research, and S.~B. Porter 
     for discussing the results of his group with us prior to publication. In preparation of this paper, we made use of the NASA 
     Astrophysics Data System, the ASTRO-PH e-print server, and the MPC data server.

  \newpage
  \appendix
  \section{Comparison between the pre- and post-New Horizons data orbital solutions}
     Table \ref{orb} shows three orbital solutions for 15810 (1994 JR$_1$). The third column corresponds to the one currently available and
     includes astrometry acquired by NASA's New Horizons spacecraft and discussed in Porter et al. (2016). As pointed out by Porter et al. 
     (2016), the New Horizons data have improved the orbital solution of 15810 very significantly. However, numerical simulations equivalent 
     to those in de la Fuente Marcos \& de la Fuente Marcos (2012c) but making use of the second and third orbital solutions in Table 
     \ref{orb} still produce the same basic results; the differences are simply too small to claim any dramatic change in the nature of the 
     orbital evolution of 15810 as a result of the new and indeed improved orbit. The original description in de la Fuente Marcos \& de la 
     Fuente Marcos (2012c) is certainly still valid. Nevertheless, the astrometry discussed in Porter et al. (2016) confirms beyond any 
     doubt the role that NASA's New Horizons spacecraft may play in improving the orbital solutions of many trans-Neptunian objects over the
     next decade or so.
%
%
%
         \begin{table*}
            \fontsize{8}{11pt}\selectfont
            \tabcolsep 0.35truecm
            \caption{Heliocentric Keplerian orbital elements of 15810 (1994 JR$_1$) from JPL's Small-Body Database and \textsc{horizons} 
                     On-Line Ephemeris System; values include the 1$\sigma$ uncertainty. The orbit in the left-hand column was the one 
                     available back in 2012 and it was used by de la Fuente Marcos \& de la Fuente Marcos (2012c); this orbit is referred to 
                     the epoch 2456200.5 JD CT (2012-September-30.0) and it was computed using 43 observations with an arc-length of 2236 d. 
                     The orbital solution in the column next to it was computed on 2015 October 05 13:55:21 \textsc{ut} and it is referred 
                     to the epoch 2457600.5 JD TDB (2016-July-31.0) TDB; it was computed using 49 observations with an arc-length of 7701 d. 
                     The third orbit is the one currently available and it was computed on 2016 June 21 15:49:21 \textsc{ut}. This new and 
                     improved orbital solution includes astrometry acquired by NASA's New Horizons spacecraft and is referred to the same 
                     epoch as the previous one; it was computed using 78 observations with an arc-length of 8002 d.}
            \begin{tabular}{lcccc}
              \hline
               Semimajor axis, $a$ (au)                          & = &  39.24$\pm$0.02     &  39.427$\pm$0.011     &  39.4224$\pm$0.0009     \\
               Eccentricity, $e$                                 & = &   0.1143$\pm$0.0003 &   0.1196$\pm$0.0002   &   0.119501$\pm$0.000010 \\
               Inclination, $i$ (\degr)                          & = &   3.8032$\pm$0.0002 &   3.80801$\pm$0.00005 &   3.80802$\pm$0.00005   \\
               Longitude of the ascending node, $\Omega$ (\degr) & = & 144.753$\pm$0.011   & 144.6859$\pm$0.0011   & 144.6854$\pm$0.0007     \\
               Argument of perihelion, $\omega$ (\degr)          & = & 102.1$\pm$0.2       & 101.55$\pm$0.03       & 101.535$\pm$0.012       \\
              \hline
            \end{tabular}
            \label{orb}
         \end{table*}
%
%

  \bsp
  \label{lastpage}
\end{document}